# High-quality electron beam generation from laser wakefield accelerators for driving compact free electron lasers

Ke Feng,[1,*,‡] Kangnan Jiang,[1,‡] Runshu Hu,[1,2] Chen Lv,[1] Xizhuan Chen,[1,2] Hai Jiang,[1] Shixia Luan,[1] Wentao Wang,[1,†] and Ruxin Li[1,2,3]

[1]State Key Laboratory of Ultra-intense laser Science and Technology, Shanghai Institute of Optics and Fine Mechanics (SIOM), Chinese Academy of Sciences (CAS), Shanghai 201800, China
[2]Center of Materials Science and Optoelectronics Engineering, University of Chinese Academy of Sciences, Beijing 101408, China.
[3]School of Physical Science and Technology, ShanghaiTech University, Shanghai 201210, China

**ABSTRACT**. Despite the successful demonstration of compact free electron lasers (FELs) driven by laser wakefield accelerators (LWFAs), the pursuit of further enhancements in high-gain compact FELs presents a challenge due to the limitations in electron beam quality. In this work, we pinpoint the pivotal physics and optimization strategies for high-quality single-stage LWFAs that are crucial for high-gain FELs. We have delved into the synergistic injection mechanism, where the self-evolution injection threshold is far from reached at the injection position, with both the shock front and self-evolution of the laser playing a role in the injection process. A thorough discussion has been provided on the beam-quality degradation and optimization strategies, in terms of global (slice) energy spread and projected (slice) emittance. With the goal of achieving high-gain FELs driven by LWFAs, we have also explored the synthesis quality of the electron beam to determine an optimized power gain length. A comprehensive start-to-end simulation has been conducted, demonstrating the effectiveness of compact FELs powered by these high-quality electron beams. The resulting radiation reaches the saturation regime after a 4.5-meter-long undulator, with an energy of 17.4 μJ and a power of 6.0 GW at a wavelength of 23.9 nm. This proposed scheme offers not only a framework for optimizing beam quality in LWFAs, but also a promising path for future compact LWFA-driven FELs to achieve saturated regimes, opening up new possibilities for widespread applications.

## I. INTRODUCTION

Laser wakefield accelerators (LWFAs) have garnered escalating attention due to their capabilities of sustaining ultra-high gradients since they were first proposed by Tajima and Dawson [1]. In LWFAs, the ponderomotive force of the laser expels plasma electrons from their equilibrium positions, and the resulting immense charge separation fields of the excited wakefield trailing behind the laser can be exploited to accelerate particles. Unencumbered by the constraints of breakdown voltage, LWFAs can achieve unprecedented accelerating gradients up to ~100 GV/m. This capability is several orders of magnitude greater than that of conventional radio-frequency (RF) accelerators, positioning LWFAs as highly promising candidates for the next generation of compact accelerators [2-4]. Over the past decade, tremendous progress has been made in LWFAs [5-12], largely driven by advanced beam and plasma diagnostics alongside the exploration of various innovative electron injection schemes that yield diverse beam characteristics [13-16]. Considerable efforts have been dedicated to improving the beam qualities, with the obtained beam charges ranging from several hundred picocoulombs to nanocoulombs [17, 18], energy spreads of a few per-mille [12, 19], emittances of the order of 0.1 mm mrad [20, 21], and with increasing stabilities [22]. Besides, several beam manipulation strategies have been explored with the aim of further enhancing the beam qualities [23-25].

Concurrent with advancements in acceleration techniques, laser plasma-based light sources, including X-ray and γ-ray sources based on betatron motion [26-28], Compton scattering processes [29-31], and compact X-ray free electron lasers (XFELs) [32, 33], among others [34]. Offering complementary alternatives to sources based on RF accelerator technologies [35], these novel approaches are being actively pursued globally. In particular, XFELs driven by LWFA-based *e*-beams are promising for generating high-brilliance X-ray lasers on a laboratory scale, and they are anticipated to enable a multitude of significant applications [36, 37]. The realization of such compact LWFA-driven FELs has been identified as one of the major challenges in this decade, as addressed in the European Plasma Research Accelerator with eXcellence In Applications (EuPRAXIA) [38]. Electron beams from LWFAs can routinely drive synchrotron undulator radiation and have been demonstrated by several research groups from visible to soft X-ray regimes [39-41]. However, realizing FEL gain continues to pose a significant challenge, primarily due to the stringent requirements on beam quality [42, 43]. Building on recent advancements in beam quality enhancement, the lasing of an FEL adopting an LWFA has been demonstrated with the self-amplified spontaneous emission (SASE) configuration at 27 nm, 40 nm and 420 nm [44-46] and the seeded configuration at 274 nm [47], respectively. Aiming at high-gain FEL to the saturated regime, improving the beam quality, in terms of 6D brightness, is of paramount importance [48, 49], which can be defined as $B_{6D} = I/(\varepsilon_x \varepsilon_y\, 0.1\%\Delta E / E)$, where $I$ is peak current, $\varepsilon_x$ and $\varepsilon_y$ are normalized transverse emittances, $\Delta E / E$ are the energy spread of the electron beam. Given the interdependence of the 6D brightness parameters, it is essential to concurrently optimize the energy spread, emittance, and beam current during the initial plasma stage of the acceleration process.

* Corresponding author: fengke@siom.ac.cn
† Corresponding author: wwt1980@siom.ac.cn
‡ These authors contributed equally to this work.

In this paper, we have identified the pivotal physics and optimization strategies crucial for the generation of high-quality *e*-beams within a single-stage LWFA. The synergistic mechanism has been investigated, where both the shock front and self-evolution of the laser play a role in the injection process [19]. Numerical simulations have revealed that the quality of the accelerated *e*-beam can be precisely controlled by fine-tuning the relative contributions of the shock-front and self-evolution of the laser during the injection process. The source of the beam quality degradation has been identified, and targeted optimization strategies have been developed, focusing on the global (slice) energy spread and the projected (slice) normalized emittance. A start-to-end simulation was conducted to demonstrate the efficacy of the derived *e*-beam in driving a compact FEL in the extreme ultraviolet (EUV) regime, with corresponding radiation energy and power reaching up to 17.4 μJ and 6.0 GW, respectively, within a 4.5-meter-long undulator. The proposed scheme offers not only a framework for optimizing beam quality in LWFAs, but also significant potential for the advancement of compact LWFA-driven FELs towards saturation, paving the way for the realization of a truly compact and widely accessible system.

## II. BEAM QUALITY OPTIMIZATION

To get insights into the beam quality optimization, spectral quasi-3D particle-in-cell (PIC) simulations were performed utilizing the FBPIC code with the co-moving simulation box [50, 51]. The simulation domain has sizes of 50 μm and 120 μm in the longitudinal and transverse directions, respectively, with the longitudinal grid size of $\Delta z$ = 31.25 nm, the transverse size of $\Delta r$ = 80 nm and 16 macro-particles per cell. A Gaussian laser pulse has a central wavelength of $\lambda$ = 800 nm, a waist radius of $w_0$ = 35 μm, a pulse duration of 25 *fs* at full width at half maximum (FWHM), and a normalized vector potential of $a_0$ = 1.3. The density profile was chosen based on the previously diagnosed fringe pattern reported in Ref. [44], as illustrated in Fig. 1(a), with the corresponding shock front lengths ranging from 50 to 100 μm. The density up-ramp at the entrance of the plasma provides a moderate distance for laser evolution. The density reaches its maximum at 2 mm from the plasma entrance, followed by a down-ramp with various lengths and a platform with a plasma density of $n_{\mathrm{plat}} = 2.18\times10^{18}\,\mathrm{cm}^{-3}$ . The density down-ramp, characterized by the gradient $\partial n/\partial z$ and its length $L$, plays a crucial role in the synergistic injection mechanism and, consequently, the acceleration. The down-ramp length, varying from 50 to 100 μm, was treated as a variable to elucidate the underlying physics involved in the parametric optimization of the beam.

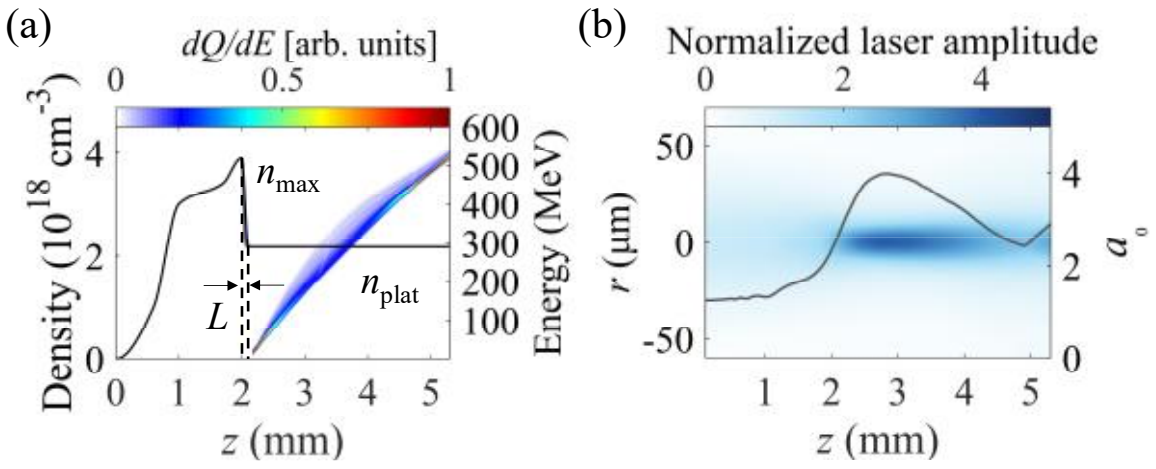

FIG. 1. Particle-in-cell simulation for high-quality LWFAs. (a) Plasma density profile and the energy evolution of the *e*-beams. The plasma density gets its maximum at the position of 2 mm, followed by a density down ramp and a plateau. (b) Evolution of the normalized laser amplitude, with the corresponding on-axis peak value represented by grey solid line. The down-ramp length was chosen to be 75 μm as an example in (a) and (b).

The energy evolution of the accelerated *e*-beam is presented in Fig. 1(a), with the corresponding down-ramp length of 75 μm. The energy spread of the *e*-beam increases during the earlier stage of acceleration, but this is subsequently compensated for, as detailed in the section "energy spread compensation" section. The beam-loading effects, describing the dependence of the wakefield on the current of the accelerated bunch, can be leveraged as an effective tool for optimizing beam parameters. The evolution of the normalized laser amplitude is depicted in Fig. 1(b). Contrary to earlier studies [52-54], the self-evolution injection threshold is far from reached prior to the density down-ramp, and both the shock-front and the self-evolution of the laser contribute to the injection process, each exhibiting distinct phenomena.

### A. Current and beam charge

Moderate beam charges, reaching up to several tens of pico-Coulombs, and short bunch durations of several femtoseconds are essential for highly demanding applications, such as compact FELs [15]. In this section, we delve into the beam current and charge behavior for the presented single-stage scheme with a range of shock front lengths. Figure 2(a) displays the beam current at 5 mm from the plasma entrance for shock front lengths varying from 50 to 100 μm. An increase in the shock front length results in a decrease in bunch duration while the peak current remains relatively stable for the situation with shock front lengths from 50 to 70 μm. In contrast, for shock front lengths ranging from 80 to 100 μm, the peak current increases while the bunch duration remains constant. Figure 2(b) illustrates the charge dependence on the shock front length. With an increase in the shock front length, the injected beam charge exhibits a trend of initial decrease followed by a subsequent increase within the presented parametric intervals.

To elucidate the beam current and charge behavior depicted in Figs. 2(a) and 2(b), an in-depth analysis of the injection process is required. Injection occurs when the phase velocity of the bubble tail drops below that of the electrons, thereby enabling the electrons to penetrate into the bubble [55]. Figures 2(c) and 2(d) illustrate the evolution of the longitudinal velocities for both the injected electrons (red solid lines) and uninjected electrons (blue solid lines), with shock front lengths of 50 and 100 μm, respectively. The threshold velocities associated with the injected electrons for the two situations are 0.9385 and 0.9554, respectively, as indicated by the black dashed lines in Figs. 2(c) and 2(d). The phase velocities of the bubble tail decrease rapidly due to the significant expansion of the bubble within the shock front, resulting in a lower injection threshold for a sharper shock front. This phenomenon explains the observed decrease in beam charge as the shock front length increases within the range of 50 to 70 μm, and such a regime can be identified as the shock front injection-dominated regime. As shown in Figs. 2(c) and 2(d), a longer shock front implies that injection occurs later. Consequently, a stronger driver intensity is expected at the injection point, as depicted in Fig. 1(b), which in turn leads to higher velocities of the plasma electrons. Despite the higher injection threshold associated with longer shock fronts, the charge of the injected electrons increases as the shock front length increases for shock front lengths ranging from 80 to 100 μm. This regime is identified as the self-evolution injection-dominated regime.

A small fraction of the electrons within the beam, those with velocities surpassing the initial injection threshold, are unable to enter the bubble, as observed in Figs. 2(c) and 2(d). At the position where injection occurs, the driver intensity is inadequate, as depicted in Figure 1(b), and the bubble undergoes distortion due to its rapid expansion. Beam-loading effects become pronounced during the injection process, causing a fraction of the initially injected electrons within the beam to encounter decelerating fields that prevent further injection into the bubble. Consequently, a secondary injection threshold is established, which is indicated by the green dashed lines in Figs. 2(c) and 2(d). For the two situations with shock front lengths of 50 and 100 μm, the secondary injection thresholds are 0.9851 and 0.9903, respectively.

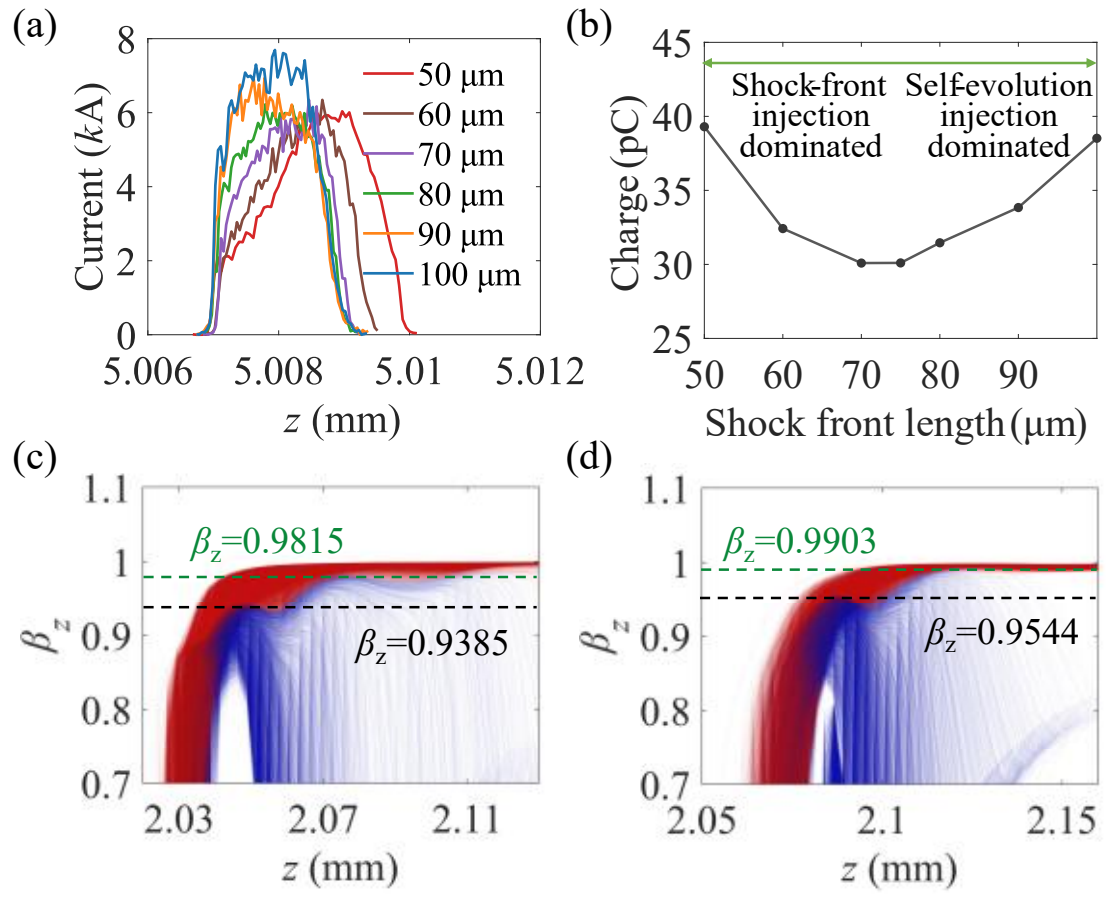

FIG. 2. (a) Beam current at 5 mm from the plasma entrance for various shock front lengths from 50 to 100 μm. (b) Injected beam charge as a function of shock front length. Evolution of the normalized velocities $\beta_z$ of the electrons for both injected (red solid lines) and uninjected (blue solid lines) with shock front lengths of (c) 50 μm and (d) 100 μm, respectively. The black and the green dashed lines denote the first and the second injection threshold velocities.

### B. Energy spread compensation

Despite the substantial progress in energy spread optimization, the attainment of sub-percent level energy spread in LWFA while preserving sufficient charge remains a significant challenge [56]. This challenge affects not only applications such as FELs which demand a relative energy spread of 0.1% level [43], but also for the beam transport after leaving the plasma, which causes dramatic chromatic emittance growth and substantially reduces the beam quality [57, 58].

The energy spread of the $e$-beam is primarily due to the inhomogeneity of the accelerating field, indicating that electrons at various positions within the beam experience different accelerating fields. Addressing the compensation of $e$-beam energy spread is fundamentally about reversing the gradient of the accelerating field at the electron positions within the beam. Significant efforts have been dedicated towards this objective, including the wake structure modulating with the laser evolution and beam-loading effects [16, 19], altering the accelerating phase with the modulated or tailored plasma profile [12], or using the external magnet chicane or plasma de-chirper for multi-stage scheme [25].

In the single-stage scheme presented, the strategy for energy spread compression is elucidated according to quasi-3D PIC simulations, as depicted in Fig. 3. The wake emerges as a nonlinear structure at early stage of the acceleration and the $e$-beam exhibits a positive chirp, attributed to the electrons at the head of the bunch experiencing lower accelerating fields compared to those at the tail (see Fig. 3(a)). As the drive laser weakens due to diffraction, the wakefield gradually evolves into a weak nonlinear structure, and the beam-loading effects become pronounced, as depicted in Fig. 3(b). Consequently, a reversed gradient of the accelerating field is imparted to the accelerated electrons, allowing for continuous compensation of the beam's chirp (as shown in Figs. 3(b) and 3(c)). The relative energy spread of the $e$-beam reaches its minimum of approximately 0.5% on FWHM at the position of 5 mm, where the chirp of the $e$-beam is fully compensated, as illustrated in Fig. 3(c). After that, the energy chirp of the $e$-beam gets overcompensated, resulting in an increasing energy spread, as depicted in Fig. 3(d). It is noted that the electrons at the tail of the bunch possess higher energy, which causes growth of the global energy spread of the $e$-beam. Nevertheless, these electrons contribute less to the compact FELs due to their relatively small proportion, as shown in Fig. 3(c).

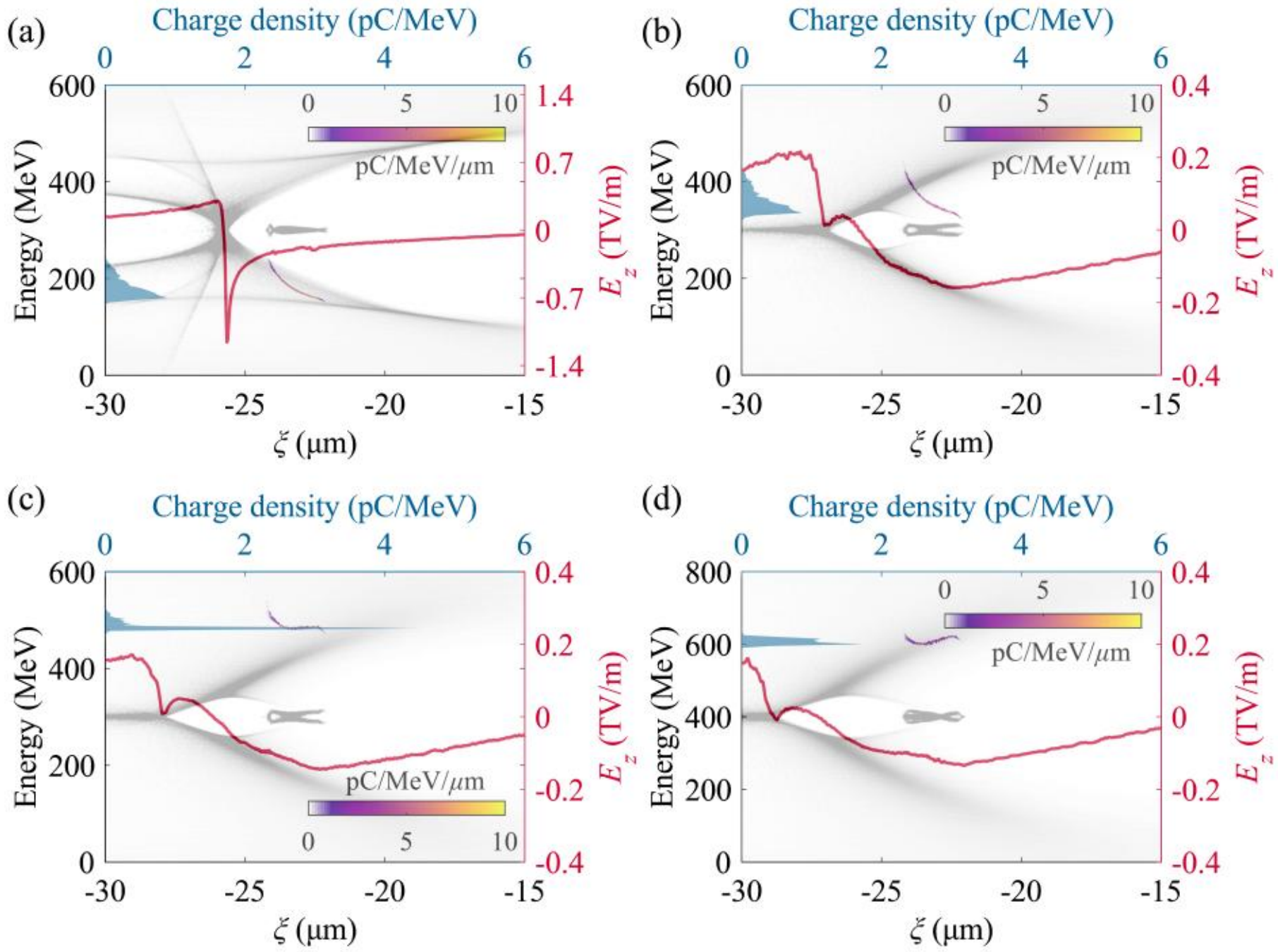

FIG. 3. Snapshots of the simulation results at the longitudinal positions of (a) 3 mm, (b) 4 mm, (c) 5 mm and (d) 6 mm, respectively, with $\xi = z - ct$. The on-axis accelerating field is represented by the red solid line, with the energy spectrum plotted as the blue shaded area on the left of each figure. The colored plot shows the phase space of the accelerated electrons, while the grey background shows the charge density map of the electrons. The shock front length of 75 μm is chosen as an example to show the mechanism of energy compensation during the acceleration process.

Figure 4 shows the evolution of the root-mean-square (RMS) energy spread $\Delta E$ and the relative energy spread $\Delta E/E$ (RES) along the acceleration distance for shock front lengths varying from 50 to 100 μm. In the early stages of the acceleration, the energy spread of the *e*-beam escalates during the acceleration due to the inhomogeneity of the accelerating field experienced by the electrons (as shown in Fig. 3(a)). The energy spread attains its maximum at the position of approximately 3.5 mm from the plasma entrance, beyond which it decreases due to the local reversal of the accelerating field gradient, as illustrated in Fig. 3. The energy spread and RES of the *e*-beam are compensated during the subsequent accelerating stage, subsequently reaching their minimum values. After that, the energy spread and RES of the *e*-beam increase as a result of over-compensation of the *e*-beam energy chirp. In the self-evolution injection dominated regime, characterized by shock front lengths ranging from 80 to 100 μm as depicted in Fig. 2(a), an *e*-beam with a higher current is attainable. Consequently, a more pronounced beam-loading effect is expected, leading to a more rapid chirp compensation process. As indicated in Fig. 4, the minimum value of the energy spread and RES are larger in the shock-front injection dominated regime, which are predominantly dictated by the slice energy spread and will be discussed in the subsequent "slice energy spread optimization" section.

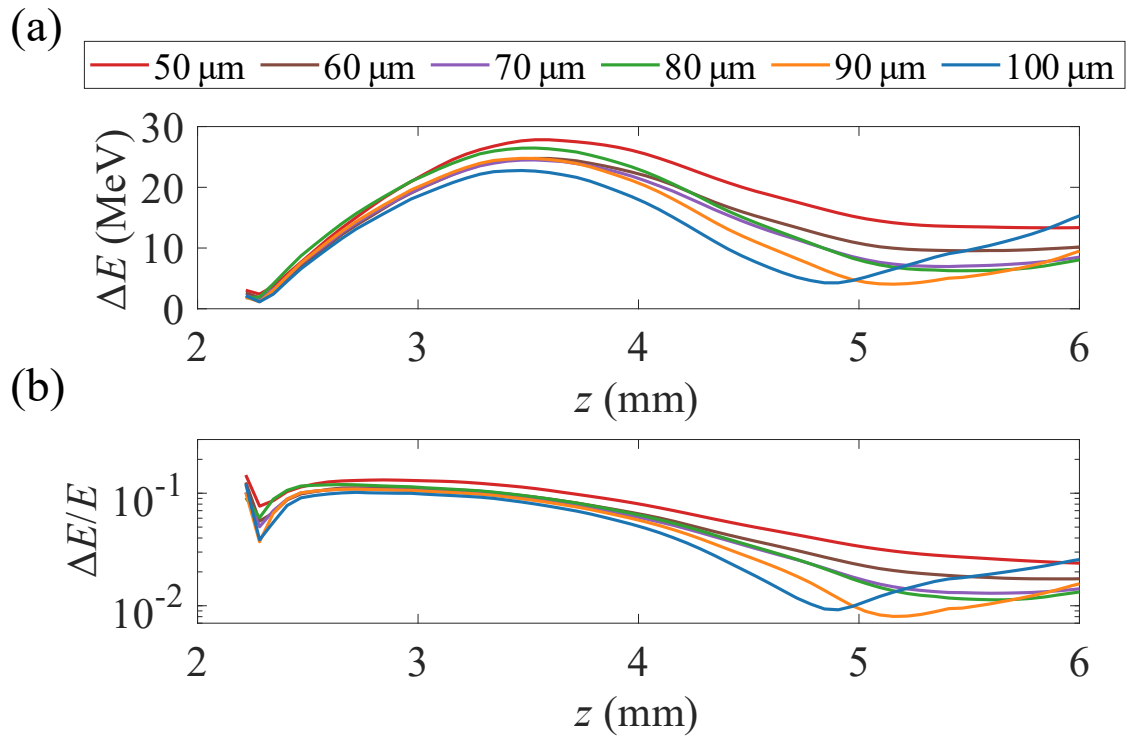

FIG. 4. (a) RMS energy spread and (b) relative energy spread (RES) of the *e*-beam along the acceleration distance for various shock front lengths from 50 to 100 μm.

## C. Slice energy spread optimization

The total energy spread of the *e*-beam originates from two aspects, the correlated and uncorrelated energy spreads. The total energy spread is predominantly determined by the correlated energy spread, which represents the aggregate spread of longitudinal slices within the bunch. The uncorrelated energy spread, known as the slice energy spread (SES), denotes the lower limit of the energy spread and becomes predominant when the chirp of the *e*-beam is fully compensated. A relative SES on the order of 0.1% could enable free electron laser (FEL) gain with plasma accelerators [43]. Various strategies, including beam decompression [59] or dispersively matched transverse gradient undulators (TGU) [60], have been proposed to alleviate the energy spread over longitudinal or energy slices. However, these approaches often come at the cost of reducing the peak current or expanding the transverse beam size. Achieving an *e*-beam with extremely low slice energy spread at the LWFA stage is of crucial importance for driving high-gain FELs.

The accelerating field experienced by a particle is the superposition of the wakefield induced by the laser and the effects of beam-loading, that is

$$E_{acc}(\xi,r,z) = E_z(\xi,r,z) + E_{zb}(\xi,r,z) \qquad (1)$$

The first term in Eq. (1) corresponds to the plasma wakefield induced by the laser, which is transversely independent near the axis for normal conditions due to the relatively small beam size compared to the driver. Consequently, the first term is independent of the radial coordinate $r$ under the paraxial approximation. The second term accounts for the beam-loading effects, which shows significant transverse dependence as elaborated in Ref. [61]. The SES is predominantly derived from two aspects, the longitudinal dependence of the accelerating field with significant particle exchange between slices [62], and the transverse difference of the accelerating field attributable to the beam-loading effects [61]. As indicated in Ref. [62], the betatron radiation can also lead to an increase in SES, which is not addressed here because it becomes dominant only in scenarios involving *e*-beams of extremely high energy.

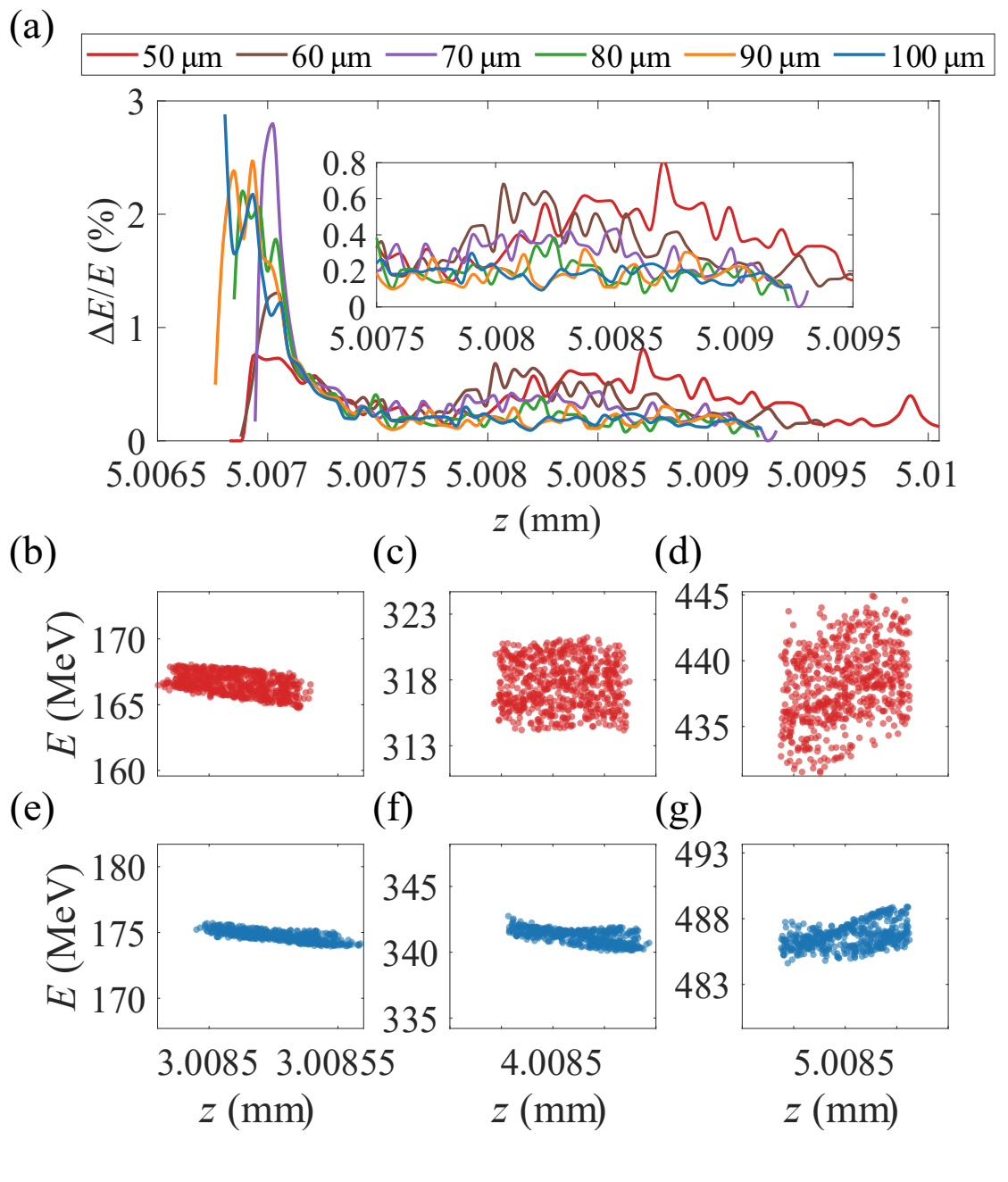

FIG. 5 (a) Relative SES of the accelerated *e*-beam at the position of 5 mm from the plasma entrance, with shock front lengths ranging from 50 to 100 μm. The local enlarged plot is also given to show the detail of the *e*-beam SES. Longitudinal phase space distributions of the selected slice near the beam center with shock front lengths of (b-d) 50 μm (red dots) and (e-g) 100 μm (blue dots), with the positions of 3 mm (b) and (e), 4 mm (c) and (f), 5 mm (d) and (g), respectively. The selected slice has a width of 50 nm.

Figure 5(a) depicts the SES along the longitudinal slices of the *e*-beam at the position of 5 mm for various shock-front lengths from 50 to 100 μm, with an accompanying local enlargement for clarity. Within the parametric range, the SES for shock-front injection-dominated (with shock-front lengths of 50~70 μm) situations is generally higher than that for self-evolution injection-dominated (with shock-front lengths of 80~100 μm) situations. In general, the average SES decreases with an increase in shock-front length under the simulation conditions presented. Specifically, an averaged SES of about 0.2% is achieved with a 100-μm-long shock front, as indicated by the blue solid line in Fig. 5(a). Although the SES increases at the tail of the bunch, it makes less contribution for driving compact FELs due to the relatively small fraction of the total bunch (see Fig. 3(c)). To elucidate the origins of the SES in the presented scheme, a typical slice near the beam center with a width of 50 nm was chosen for tracking. The longitudinal phase space distributions at various positions for the selected slice are illustrated in Figs. 5(b)−5(g), with the shock front lengths of 50 and 100 μm, respectively. The particles are observed to remain predominantly within the same longitudinal range, indicating that electron slippage and particle exchange between slices are negligible during the acceleration process. Consequently, the growth of the SES is not attributable to the transverse oscillation of the electrons (betatron motion).

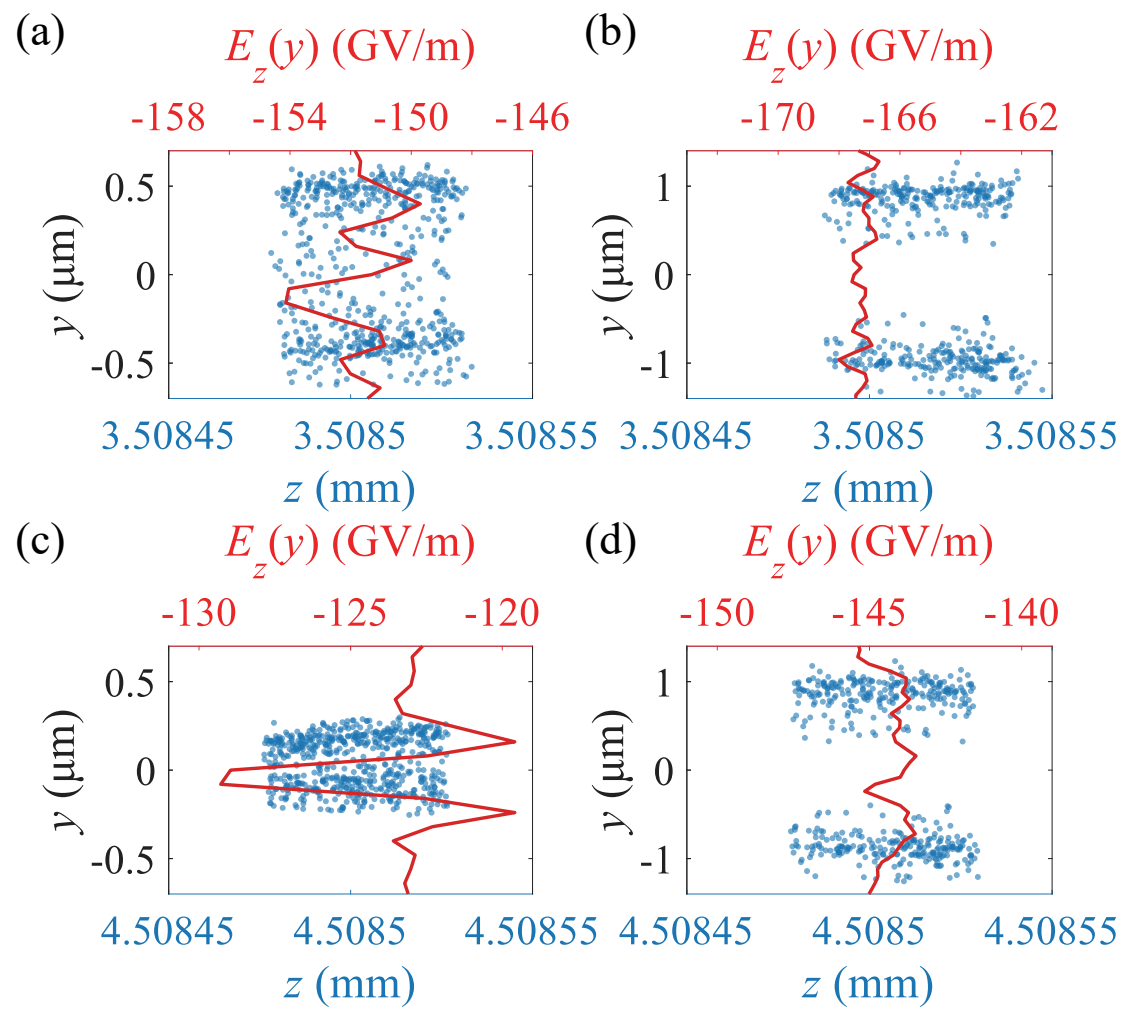

FIG. 6. Spatial (*z, y*) distributions of the selected electron in one selected slice near the beam center and the corresponding field experienced by the electrons along the vertical directions at the positions of (a) and (b) 3.5 mm, (c) and (d) 4.5 mm, with the shock front length of (a) and (c) 50 μm, (b) and (d) 100 μm, respectively.

Figure 5 illustrates that the SES exhibits more rapid growth with a steeper shock front. Such disparity in SES behavior might stem from the transverse dependence of the accelerating field. Consequently, an analysis of the acceleration field experienced by electrons at various transverse positions was conducted. Fig. 6 presents the spatial (*z, y*) distributions of the electrons within the selected slice, alongside the corresponding accelerating field along the transverse direction. The transverse inhomogeneity of the accelerating field suggests a heightened correlation between the accelerating field and the transverse position under the condition with a steeper shock front of 50 μm, as depicted in Figs. 6(a) and 6(c), leading to the significant increase of the SES (see Fig. 5). As elaborated in the emittance optimization section, the transverse emittance of the *e*-beam is substantially affected by the driver intensity and the defocusing field experienced by the electrons during the injection process. A smaller transverse emittance is obtained for the 50-μm-long shock front situation, indicating a higher transverse density of the *e*-beam and a more significant transverse dependence of the acceleration field due to the beam loading effects. This transverse dependence is a key factor contributing to the growth of SES of the *e*-beam with a small betatron amplitude. This transverse dependence can be mitigated by employing a stronger driver to effectively expel the electrons and create a clearer ion background in the bubble [61], or at the expense of transverse emittance of the *e*-beam to weaken the transverse dependence of the accelerating field (as shown in Fig. 5 and Fig. 7).

### D. Projected emittance optimization

The generation and acceleration of the *e*-beam with sufficiently low emittance is crucial for a range of prominent accelerator applications, particularly in high-energy physics and advanced light sources [63, 64]. In the realm of high-energy physics experiments, emittance is a critical factor that limits the focus-ability of particle beams, thereby influencing the achievable luminosity and consequently, the event rate [65]. Obtaining *e*-beams with ultra-low emittance to reach the luminosity goals in colliders could potentially open new frontiers in high-energy physics research. Similarly, the low emittance is imperative for light sources based on inverse Compton scattering, leading to a high photon flux due to the small spot sizes during the scattering process [29-31]. In addition to the energy spread, a sufficiently low emittance of the *e*-beam is significant for FELs, as the electrons with varying betatron amplitudes spread out in longitudinal positions and phases relative to the radiation with a finite-emittance *e*-beam and hence lead to degradation of FEL gain [42].

Emittance is a measure of the area in the transverse phase space occupied by a beam of particles, and it can be defined by $\varepsilon_x = \gamma_0 \beta_0 \sqrt{\langle x^2 \rangle \langle x'^2 \rangle - \langle xx' \rangle^2}$ in the horizontal direction (with an equivalent definition in the vertical direction), where $\beta_0$ and $\gamma_0$ are the relativistic velocity and the Lorentz factor of a reference particle, respectively. During acceleration, electrons perform transverse oscillations at the betatron frequency $\omega_\beta = ck_\beta = ck_p / \sqrt{2\gamma}$, where $k_\beta$ is the wave number of the betatron oscillation, $k_p$ is the plasma wave number and $\gamma$ is the relativistic factor of the particle [66]. Due to the energy dependence of the betatron frequency, distinct energy slices within the beam rotate at varying rates, causing the beam to spread out in the transverse phase space. This chromatic dephasing results in an increase in the emittance of the entire bunch, referred to as the projected emittance [67]. The projected emittance serves as a general indicator of the transverse beam quality and needs to be optimized. Figure 7 illustrates the evolution of the transverse projected emittance throughout the acceleration stage for shock front lengths ranging from 50 to 100 μm. Generally, the projected emittance exhibits a small range of fluctuations during the acceleration stage, primarily induced by the chromatic dephasing and rephasing processes. The projected emittance is predominantly determined by the intrinsic emittance of the *e*-beam when it is born (as shown in Fig. 7). In the shock-front injection-dominated situation with steeper shock fronts, the projected emittance of the *e*-beam is lower compared to the self-evolution injection-dominated situation. It is noteworthy that the projected emittance is larger in the vertical plane due to the polarization of the main drive laser.

The situations with shock front lengths of 50 and 100 μm are taken as examples to explore the source and distribution of the projected emittance. Figure 8 presents the evolution of the transverse emittance and the phase space snapshots of the accelerated *e*-beam at the beginning of the acceleration. The vertical emittance of the *e*-beam is typically larger than that in the horizontal direction due to the polarization of the main drive laser. Nevertheless, the projected emittance exhibits similar evolutionary trends in both the horizontal and vertical directions. Only the horizontal phase spaces are discussed in Fig. 8 for brevity. The projected emittance is relatively low when the electrons are first injected with the shock-front length of 50 μm, but it rapidly increases from approximately 33 nm rad to over 100 nm rad in the horizontal direction within a very short distance of 100 μm, as depicted in Figure 8(a). This significant increase is primarily attributed to betatron decoherence, as evidenced in Figures 8(c) to 8(e). Initially, electrons from different slices occupy a small range in phase space after injection, but they subsequently span over $\Delta\phi \geq \pi$ in the transverse phase spaces, leading to full decoherence and maximizing the occupied phase space area, where $\Delta\phi$ is the phase differences between the first and last injected slices. However, the occupied phase space remains confined to a small range and the projected emittance exhibits only a modest increase in the self-evolution injection-dominated situation with the shock-front length of 100 μm, as shown in Figs. 8(f)−8(h). Despite full decoherence, the projected emittance remains smaller in the situation with the 50-μm-long shock front, which is attributed to the smaller emittance over slices and will be discussed in the "slice emittance optimization" section. It is noted that an extremely small phase difference is attainable only with an extremely short injection duration, which corresponds with a small beam charge.

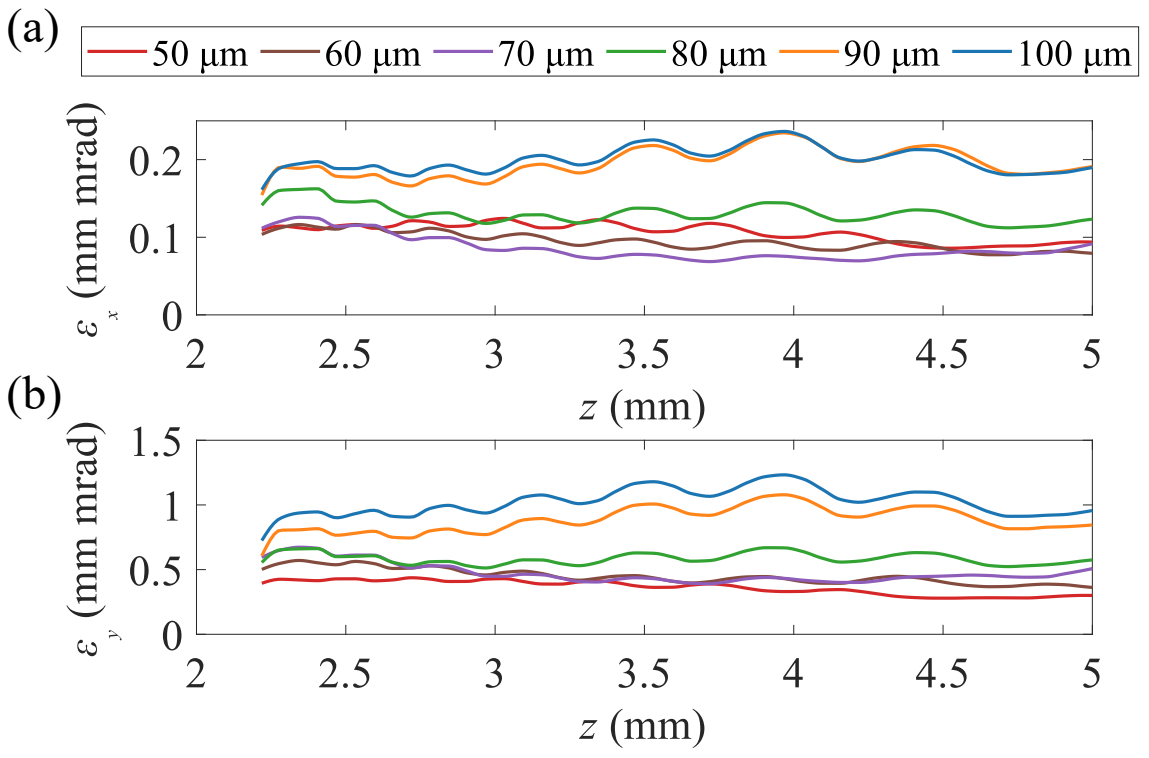

FIG. 7. Evolution of the transverse projected emittance during the acceleration stage in (a) horizontal and (b) vertical directions, respectively. The corresponding lengths of the shock front are 50 μm (red), 60 μm (brown), 70 μm (purple), 80 μm (green), 90 μm (orange), and 100 μm (blue).

As shown in Fig. 9, the difference in projected emittance evolution between the two cases arises primarily from the energy-dependent betatron oscillation. In the shock-front injection-dominated regime, the electron bunch carries a large energy chirp. Since the betatron frequency depends directly on the electron energy $\gamma$ , this large chirp results in markedly different oscillation frequencies for electrons at different longitudinal positions within the bunch. This energy-dependent frequency spread causes rapid and complete decoherence of different slices in the transverse phase space, leading to a fast growth of the projected emittance, as illustrated in Figs. 9(a) and 9(b). In contrast, for a shock-front length of 100 μm, the energy chirp is smaller and the variation in betatron frequency among slices is reduced. Consequently, the decoherence proceeds more slowly, and the slices have not yet fully decohered in the transverse phase space, as shown in Figs. 9(c) and 9(d).

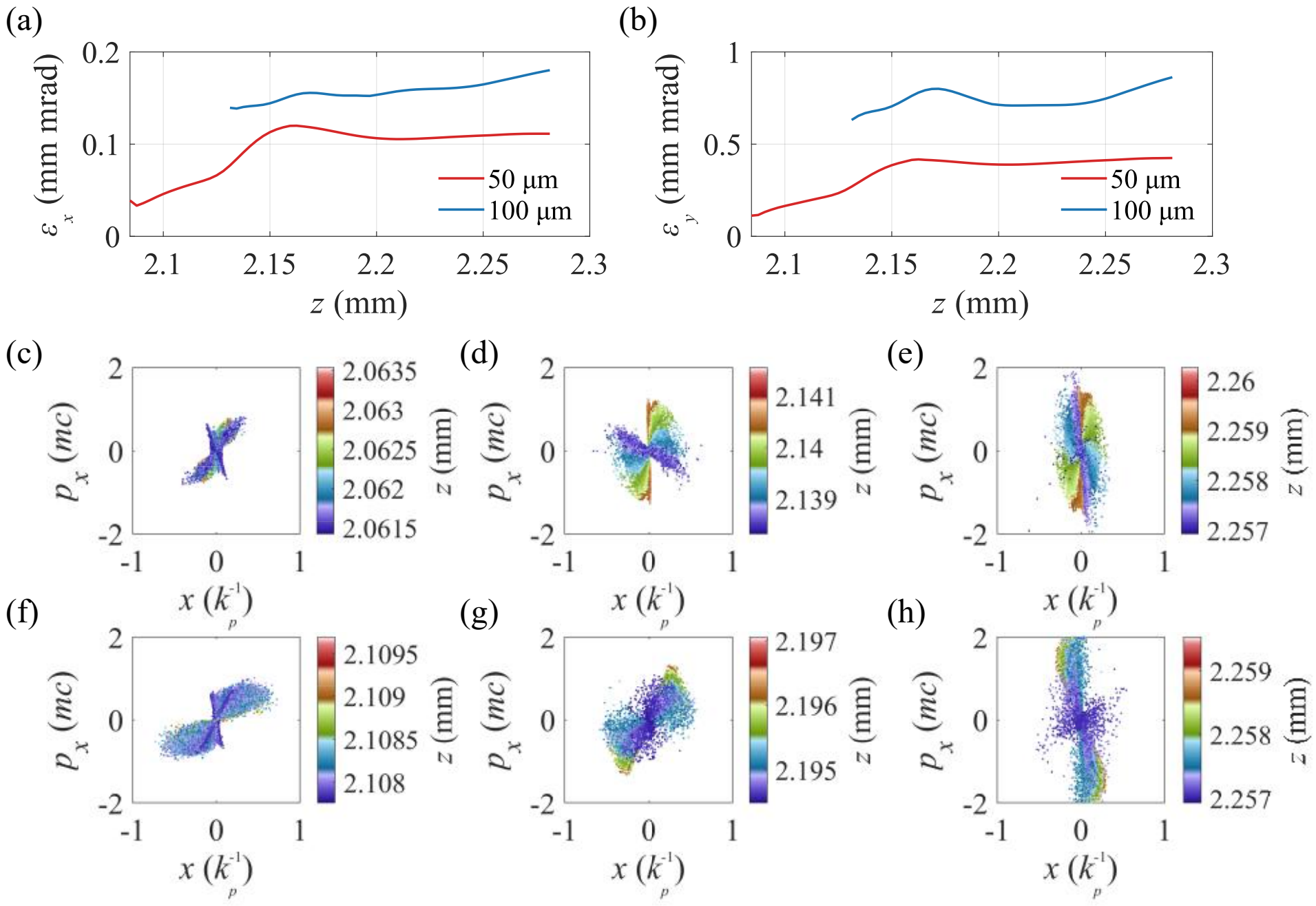

FIG. 8. Evolution of the transverse projected emittance at the beginning stage of the acceleration in (a) horizontal and (b) vertical directions, with the shock front lengths of 50 μm (red) and 100 μm (blue), respectively. (c−h) Evolution of $x-p_x$ phase spaces of the *e*-beam at different positions with shock front lengths of (c−e) 50 μm and (f−h) 100 μm, with colors representing the longitudinal positions.

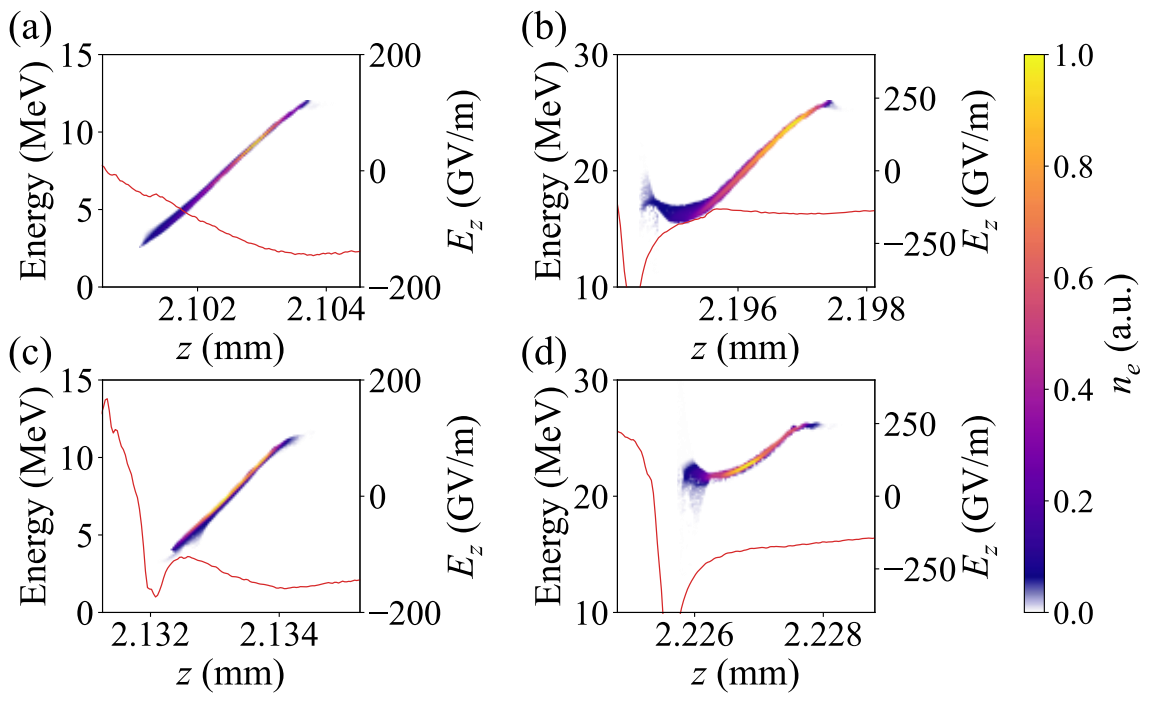

FIG. 9. Longitudinal phase-space ($z$, $E$) distributions of electrons at the beginning of the acceleration stage for shock-front lengths of 50 μm (a), (b) and 100 μm (c), (d), where $E$ denotes the electron energy. The on-axis longitudinal accelerating field $E_z$ is shown as red solid lines. The longitudinal positions correspond to those between the locations shown in Figs. 8(c)–8(e) for the 50-μm case, and to those between the locations shown in Figs. 8(f)–8(h) for the 100-μm case, respectively.

### E. Slice emittance optimization

The projected emittance can be considered as a composite of chromatic emittance and slice emittance. Chromatic emittance is characterized by the degree of decoherence over temporal slices in phase space, which may arise from phase angle mismatch or the beam size mismatch, as discussed in the preceding section. The emittance associated with each individual slice is termed slice emittance. Several compensation or matching strategies have been proposed to align each temporal slice in phase space and the projected emittance can be optimized down to its slice emittance [68, 69]. Consequently, the slice emittance determines the best achievable beam quality. In an FEL, coherence develops locally over short fractions of the whole bunch, leading to localized areas where lasing is initiated. The slice beam quality over a slippage length is of interest and significantly determines the FEL's performance [48].

Figure 10 illustrates the evolution of the average slice emittance throughout the acceleration, with the slice length equal to the longitudinal grid size in the simulation. As indicated in Fig. 10, the average slice emittances exhibit minor fluctuations during acceleration, a result of the linear radial focusing effect exerted by the laser-induced wakefield [70]. The average slice emittance of the *e*-beam is predominantly determined by the injection process where the electrons are initially born. Notably, the average slice emittance is found to be lower in the shock-front injection-dominated regime with a steeper density down-ramp (see Fig. 10). The average slice emittance at 5 mm from the plasma entrance is 0.036 (0.106) and 0.129 (0.504) mm mrad in the horizontal (vertical) directions for the situations with shock front length of 50 and 100 μm, respectively. It is noted that a larger slice emittance is anticipated in the vertical direction caused by the polarization of the driver.

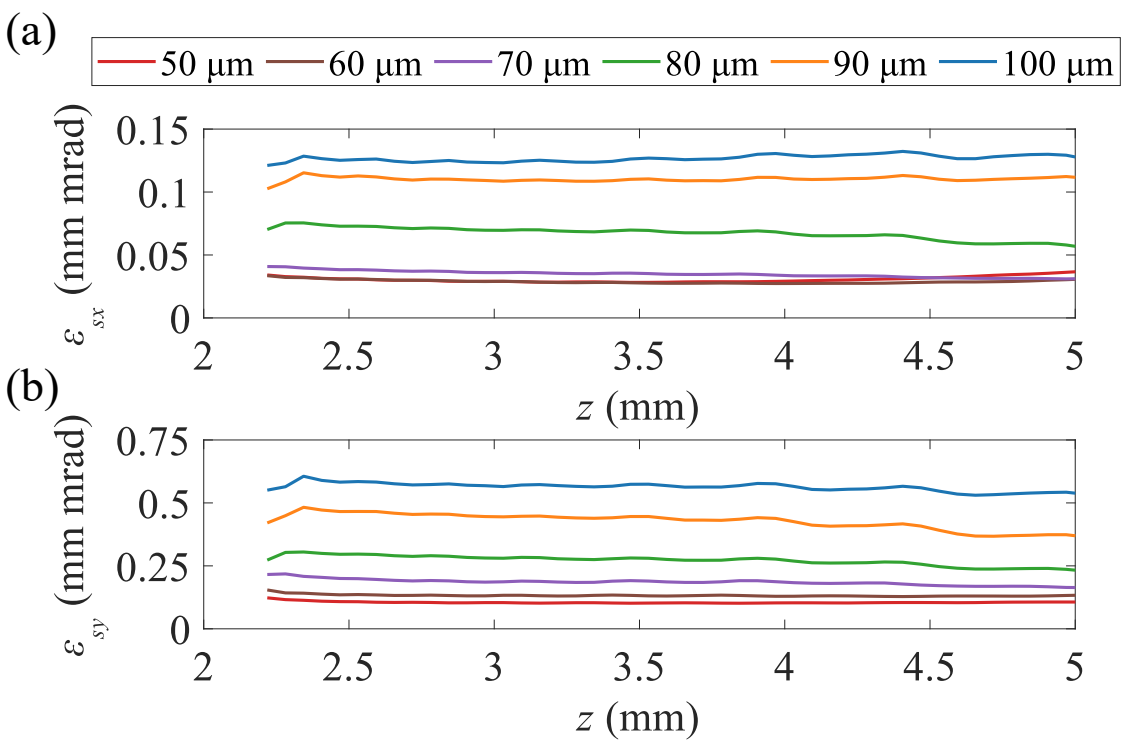

FIG. 10. Evolution of the average slice emittance during the acceleration with various shock-front lengths in (a) horizontal and (b) vertical directions, respectively. The corresponding shock-front lengths are 50 μm (red), 60 μm (brown), 70 μm (purple), 80 μm (green), 90 μm (orange), and 100 μm (blue).

As indicated in Fig. 10, the average slice emittance is primarily determined by the injection process. The situation with shock front lengths of 50 and 100 μm are taken as examples to illustrate the behaviors of the slice emittance. As pointed out in Ref. [53], the trajectories of the electrons in the blowout regime vary significantly depending on their initial positions and the driver intensity. The electrons are tracked since they are initially born, and the dependence of the injected electrons on their initial transverse positions in *x* (horizontal) and *y* (vertical) directions is presented in Figs. 11(a) and 11(b), with the shock front lengths of 50 and 100 μm, respectively. It is noted that the laser propagates along the *z* (longitudinal) direction. A larger average initial position of the electrons is expected for the 100-μm-long shock front situation, and it is more pronounced in the polarized direction of the drive laser. The injected electrons display a hollow profile in the polarized direction, with a reduced electron density at the axis. Figure 11(c) illustrates the evolution of the normalized laser amplitude near the injection position. As shown in Figs. 2(c) and 2(d), injection occurs at the end region of the shock front, indicating a longer distance for laser evolution and a higher driver intensity for the situation with the shock front length of 100 μm (as shown in Fig. 11(c)). Such difference of the driver intensity at the injection position leads to distinct behaviors of the injected charge, which are dependent on the initial positions of the electrons. The injected electrons with a stronger driver tend to have a larger average initial off-axis position.

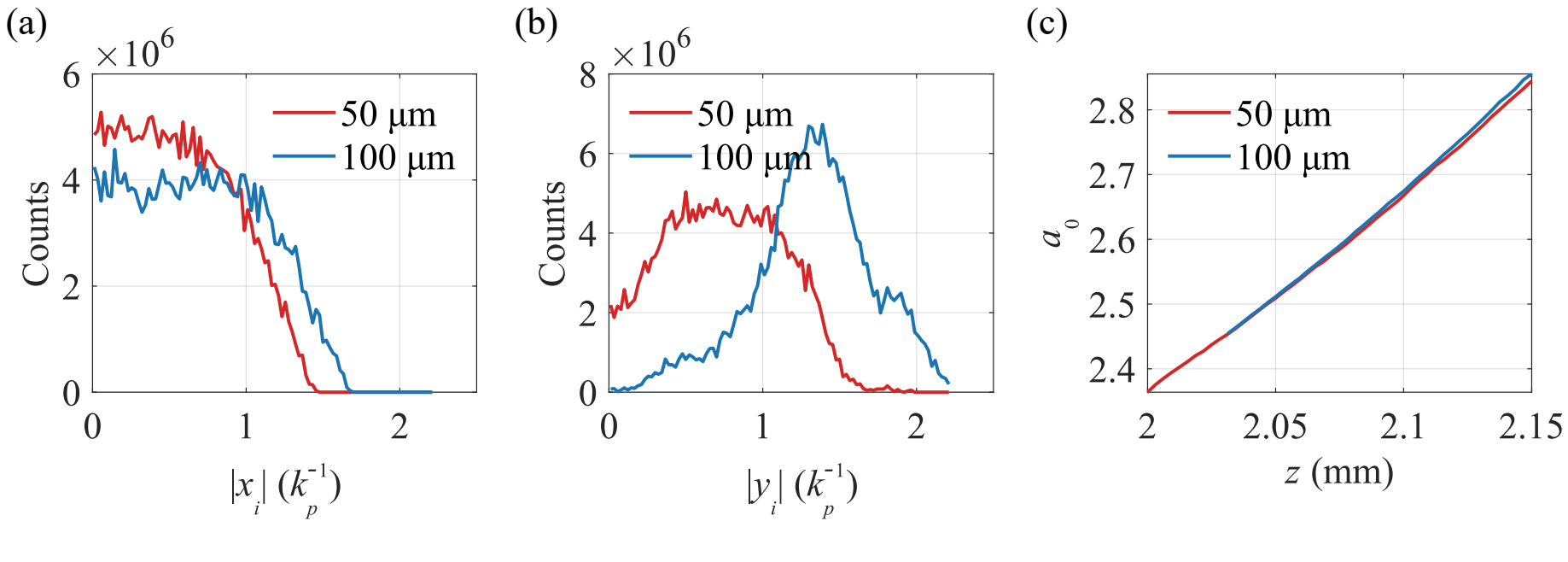

FIG. 11. Counts of the injected electrons as a function of their initial (a) horizontal $|x_i|$ and (b) vertical $|y_i|$ positions, with shock front lengths of 50 (red solid lines) and 100 (blue solid lines), respectively. The counts denote the number of injected electrons. (c) Evolution of the normalized laser amplitude near the injection position, with shock front lengths of 50 (red solid line) and 100 μm (blue solid line).

Figure 12 illustrates the transverse phase space trajectories ($x$, $p_x$) and the corresponding transverse fields experienced by the injected electrons in the selected slice. The shock front lengths are 50 μm for Figs. 12(a)−12(c) and 100 μm for Figs. 12(d)−12(f). The trajectories with different colors represent the different initial transverse positions of the injected electrons. Generally, the injected electrons with smaller initial $x_i$ tend to rotate with a small ($x$, $p_x$) in transverse phase spaces, as depicted in Figs. 12(a) and 12(d), indicating a small slice emittance. A typical trajectory for an injected electron is also plotted in the black solid line in Figs. 12(a) and 12(d), with the arrow representing the direction of motion. Initially, electrons with nearly zero transverse momentum are expelled from their initial positions and pushed away from the axis due to the ponderomotive force of the laser. Subsequently, the restoring forces induced by the wakefield draw the electrons back towards the axis, increasing their transverse momentum, as shown in Figures 12(a) and 12(d). As the wakefield rapidly expands during the shock front, the bubble gets distorted and beam-loading effects become significant. The electrons encounter a defocusing field, which decelerates them transversely as they move towards the axis, thereby reducing the transverse emittance. After that, the electrons rotate in the transverse phase space under the restoring force induced by the wakefield and are injected into the bubble. The defocusing field experienced by the electron during the injection process contributes to the compensation of the slice emittance. Figs. 12(b), 12(c), 12(e) and 12(f) illustrate the forces experienced by the injected electrons with shock front length of 50 μm (Figs. 12(b) and 12(c)) and 100 μm (Figs. 12(e) and 12(f)), with positive values denote focusing forces. Notably, the defocusing forces experienced by the injected electrons are stronger in the case of a 50 μm shock front, leading to more effective slice emittance compensation. In summary, the slice emittance can be optimized with a steeper shock front within the discussed parametric regime, which benefits from a lower driver intensity at the injection position and a better compensation process.

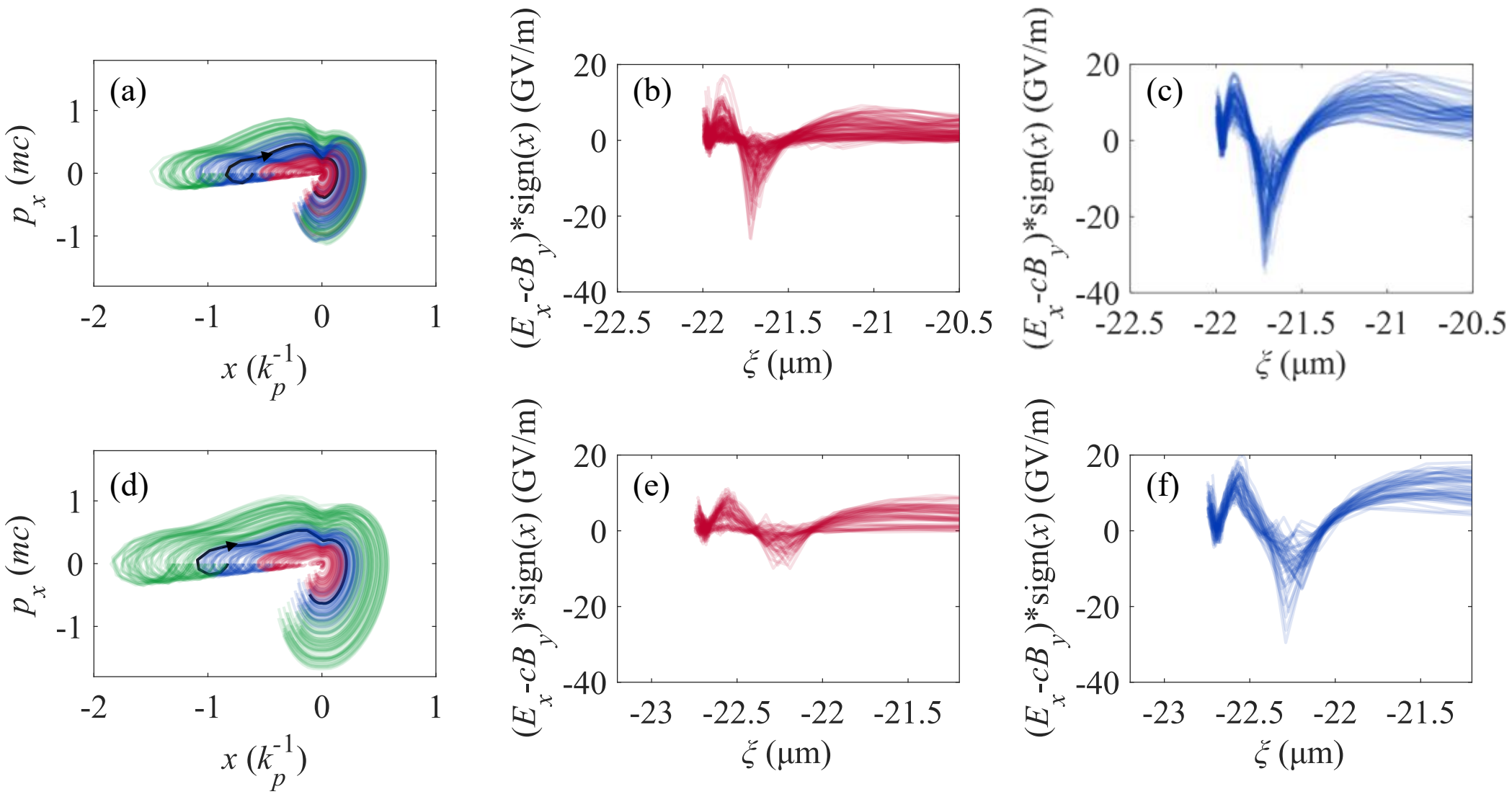

FIG. 12. Transverse phase space ($x$, $p_x$) trajectories of the injected electrons and the associate field experienced by the electrons with shock front lengths of (a−c) 50 μm and (d−f) 100 μm. The spatial coordinates are normalized with $k_p^{-1}$. The trajectories of the electrons and the field experienced by the electrons with initial positions of $-0.4k_p^{-1} < x_i < 0$, $-0.8k_p^{-1} < x_i < -0.4k_p^{-1}$ and $x_i < -0.8k_p^{-1}$ are represented by red, blue and green solid lines. The black solid line in a) and d) denotes a typical trajectory of an injected electron, with the arrow representing the direction of the motion. The sign of the transverse position is multiplied on the transverse field $E_x - cB_y$, with a positive value denoting the focusing field. The electrons are chosen from one slice in the beam center with a slice width of the simulated grid size of 31.25 nm.

## III. COMPACT FELS DRIVEN BY LWFAS

Based on the comprehensive analysis above, the trade-off between energy spread and emittance necessitates a balanced optimization. For high-gain

FELs driven by LWFAs, the *e*-beam brightness is the crucial figure of merit which amalgamates the trinity of longitudinal phase space density (energy spread $\Delta E/E$), the transverse phase space density (normalized emittance $\varepsilon_x/\varepsilon_y$) and the longitudinal charge density (current *I*) [71]. The five-dimensional (5D) and six-dimensional (6D) brightness of the *e*-beam are defined as $B_{5D}=I/(\varepsilon_x\varepsilon_y)$, $B_{6D}=I/(\varepsilon_x\varepsilon_y\, 0.1\%\Delta E/E)$, respectively, and are depicted in Fig. 13(a). With the presented single-stage LWFA configuration, the 5D- and 6D-brightness of the *e*-beam over slices typically falls within the range of $10^{17}$~$10^{18}$ (A/m$^2$, A/m$^2$/0.1%BW). Generally, the beam brightness, encompassing both 5D- and 6D-brightness, decreases with increasing shock front length, primarily due to the significant expansion of the transverse emittance of the *e*-beam (as shown in Fig. 10). It should be noted that the *e*-beam brightness is evaluated at the end of the acceleration within the plasma in Fig. 13(a). However, after extracting from the plasma, the *e*-beam parameters undergo substantial changes. Figure 13(b) presents the 5D- and 6D-brightness of the *e*-beam after its transition from plasma to vacuum, with the corresponding density profile selected based on matching conditions derived from previously reported models [72]. Notably, the 6D-brightness of the *e*-beam peaks at a shock front length of 75 μm, offering valuable insights for future optimization of the system.

To achieve the optimal conditions for driving compact FELs, the 3D power gain length $L_G$ needs to be optimized, which is defined as $L_G=L_{G0}(1+\Lambda)$, where $L_{G0}=\lambda_u/(4\sqrt{3}\pi\rho)$ is the 1D power gain length of a monoenergetic beam, $\lambda_u$ is the period length of the undulator, $\rho$ is the dimensionless Pierce parameter, $\Lambda$ is the gain length degradation factor and can be written as [42, 48]:

$$\begin{aligned}\Lambda &= a_1\eta_d^{a_2}+a_3\eta_\varepsilon^{a_4}+a_5\eta_\gamma^{a_6}+a_7\eta_\varepsilon^{a_8}\eta_\gamma^{a_9}\\ &+a_{10}\eta_d^{a_{11}}\eta_\gamma^{a_{12}}+a_{13}\eta_d^{a_{14}}\eta_\varepsilon^{a_{15}}+a_{16}\eta_d^{a_{17}}\eta_\varepsilon^{a_{18}}\eta_\gamma^{a_{19}}.\end{aligned} \tag{2}$$

Equation (2) is a fitting formula derived from Ming Xie, which depends on three scaled parameters that represent the diffraction, angular spread and energy spread, and can be written as:

$$\eta_d=\frac{L_{G0}\lambda_1}{4\pi\sigma_r^2}$$
$$\eta_\varepsilon=\frac{4\pi L_{G0}\varepsilon}{\beta_{av}\lambda_1} \tag{3}$$
$$\eta_\gamma=\frac{4\pi L_{G0}\sigma_\delta}{\lambda_u}$$

where $\lambda_1$ is the radiation wavelength, $\sigma_r$ is the average *e*-beam size throughout the undulator, $\beta_{av}$ is the average Twiss parameter in the undulator, $\sigma_\delta$ is the relative RMS energy spread and $a_1\sim a_{19}$ are the fitting coefficients without physical meaning. Minimizing the three scaled parameters leads to a minimizing $\Lambda$, and thus an ideal $L_G$. Each of the three scaled parameters is contingent upon the *e*-beam parameters after extraction from the plasma into vacuum and can be applied to evaluate the critical parameters that constrain the FEL gain. As depicted in Fig. 13(c), $\eta_d$ and $\eta_\varepsilon$ exhibit opposite trends with an increasing shock front length, indicating contrasting requirements for beam emittance [48]. $\eta_\gamma$ is still the primary issue for FEL gain degradation, which initially decreases and then stabilizes as the shock front length increases. An average Twiss parameter $\beta_{av}=10$ m is assumed, and approximately 90% of the electrons in the beam head are considered when evaluating the three scaled parameters. Figure 13(d) illustrates the 1D and 3D power gain lengths. As the shock front length increases, the 1D power gain length exhibits a monotonically increasing trend, while the 3D power gain length initially decreases and then increases. The 3D power gain length reaches its minimum with the shock front length of 75 μm, and the electrons are extracted from the plasma, demonstrating the optimal conditions for driving a compact FEL.

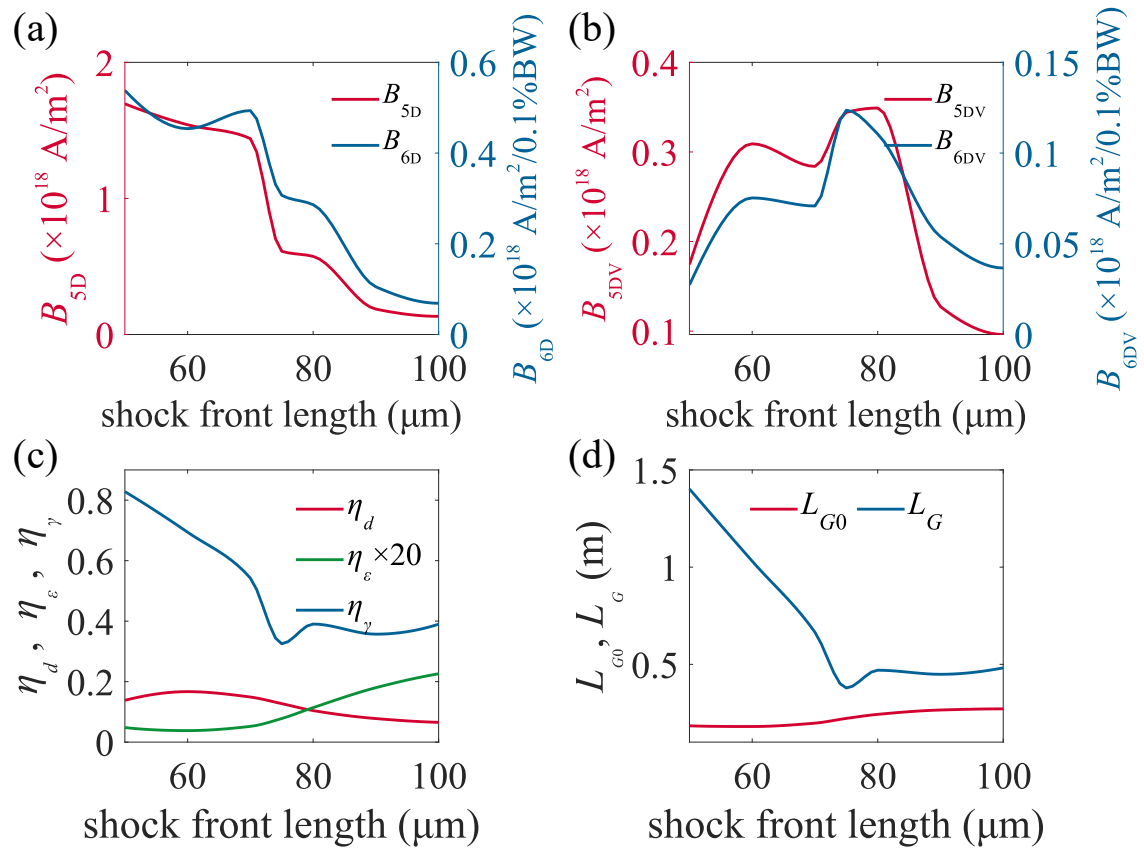

FIG. 13. Averaged 5D and 6D brightness over slices (a) at the end of the acceleration and (b) after extracting from the plasma to vacuum. (c) The three scaled parameters of diffraction $\eta_d$, angular spread $\eta_\varepsilon$ and energy spread $\eta_\gamma$ given in Equation (3). (d) 1D and 3D power gain length $L_{G0}$ and $L_G$ as a function of the shock front length. It is noted that approximately 90% of the electrons in the beam head are considered

when evaluating beam parameters and the whole beam was divided into 64 slices.

Fig. 14(a) presents the density profile with the shock front length of 75 μm, along with the corresponding energy evolution of the accelerated *e*-beam. The density platform extends to a position of 5.3 mm from the plasma entrance, where the energy chirp of the *e*-beam gets almost compensated. Following the density platform is the phase space matching stage, during which the electrons are extracted from the plasma under matching conditions consistent with previously reported models [72]. Figure 14(b) depicts the layout of the beamline, the evolution of the beam envelope and the normalized transverse emittance of the *e*-beam. The initial *e*-beam used for tracking within the beamline was obtained from FBPIC code. The beamline is equipped with two permanent quadrupoles with magnetic field gradients of 250 T/m and three electromagnetic quadrupoles with adjustable gradients ranging from 0 to 80 T/m (represented by the red squares in Fig. 14(b), consistent with our previous report [44]). The presented beamline was optimized to minimize the averaged beta function of the *e*-beam along the undulator, with the focal position located at the undulator center. The optimization variables include the spacing between the two permanent quadrupoles and the gradients of the three electromagnetic quadrupoles, using the Ocelot code [73], with space-charge and second-order effects considered. The normalized transverse emittance experiences an increase from 0.19/0.75 to 0.29/0.98 mm mrad in horizontal and vertical directions, respectively, attributable to chromatic issues. It should be noted that the projected emittance of the *e*-beam at the beginning of the beamline is larger than the values presented in Figures 5(a) and 5(c), which arises from the degradation that occurs during the extraction stage from plasma to vacuum. In addition, Bayesian optimization offers an alternative approach for such beamline optimization, as demonstrated in Ref. [74].

After being focused by a series of quadrupoles, the *e*-beam was sent to the undulator with a period length of 2.5 cm and 180 periods. The FEL interaction was simulated using three-dimensional time-dependent code GENESIS [75], employing the start-to-end model [76]. The FEL achieves saturation in a 4.5-meter-long undulator in the EUV regime, with a central wavelength of 23.9 nm, as shown in Figs. 14(c)-14(e). The corresponding radiation energy and power can reach 17.4 $\mu$J and 6.0 GW, respectively, averaged over 20 simulations with distinct random seeds.

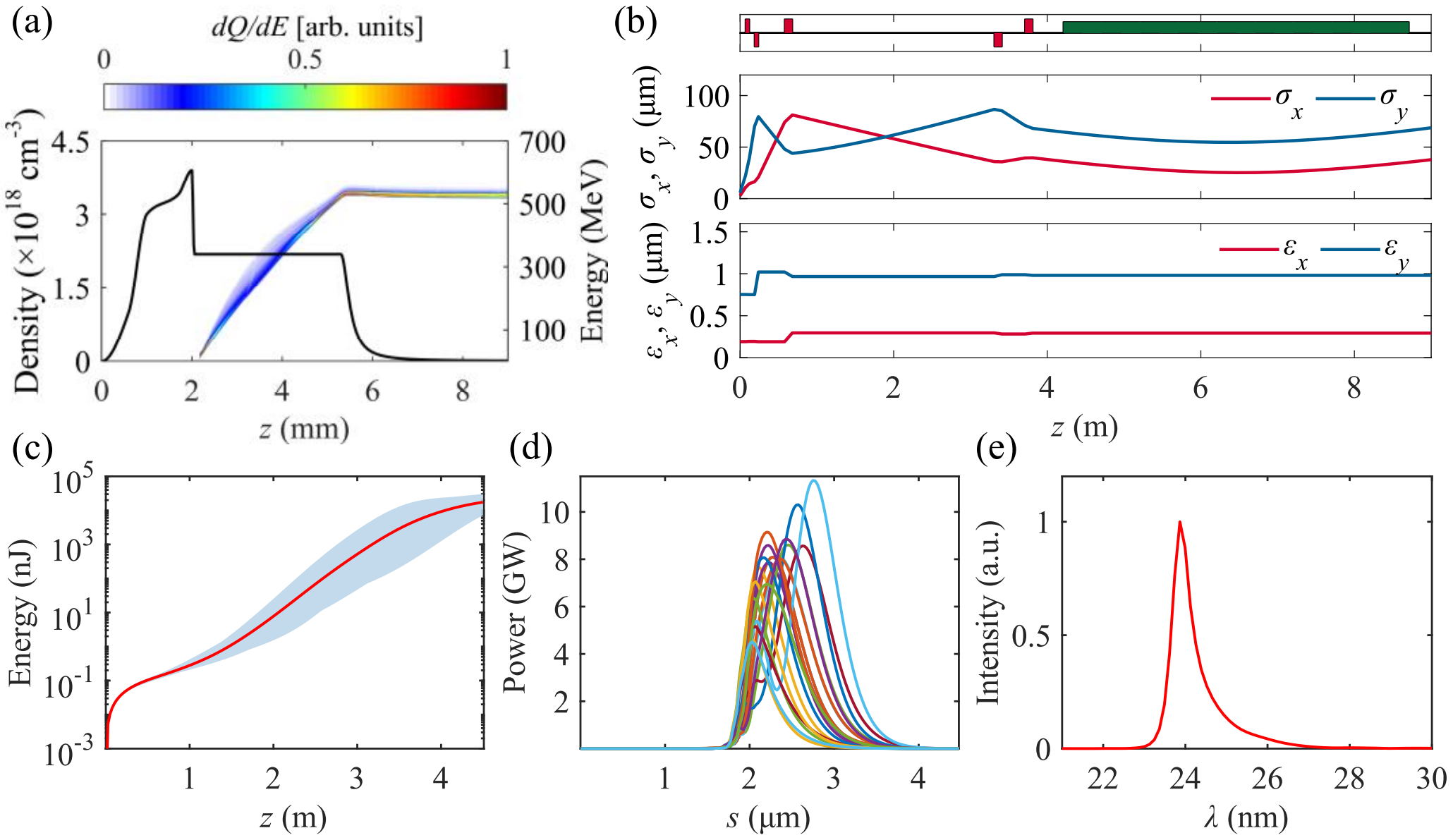

FIG. 14. (a) The density profile (black solid line) used in the simulation with the shock front length of 75 μm and the evolution of the energy spectrum d$Q$/d$E$ as a function of longitudinal position, where $Q$ and $E$ are the beam charge and energy, respectively. (b) Evolution of the beam size envelope, the normalized beam emittance along the beamline in horizontal (red solid lines) and vertical (blue solid lines) directions, with quadrupoles and undulators represented by red and green squares, respectively. (c) Radiation energy along the undulator distance. (d) Pulse duration of 20

runs with distinct random seeds and (e) spectrum of the produced coherent photon pulses at the exit of the undulator. The red solid line corresponds to the average value of the 20 runs in (c) and (e).

## IV. DISCUSSIONS

The start-to-end simulations presented above highlight the inherent potential of the proposed scheme to generate high-brightness electron beams and, consequently, to drive compact free-electron lasers (FELs). In practical experiments, however, the drive laser is subject to various fluctuations—most notably, jitter in pulse energy and focal-position shifts caused by wavefront distortion—which can degrade beam quality and undermine the stability required for user operations. Here, we briefly discuss the influence of initial laser parameters on the resulting electron beams.

The laser parameters in our scenario are based on a self-developed 200-TW laser system. Measurements over 5,400 consecutive shots within 90 minutes yielded an RMS energy fluctuation of 0.54% [77]. Assuming energy jitter within ±1.0% (approximately ±2 times the RMS), the resulting energy and charge jitter of the accelerated electron beams are within ±2.0% and 24.4%, respectively. Separately, a focal-position deviation of 0.2 mm induced by wavefront distortion has been reported [22]. With a focal jitter of ±0.4 mm (±2 times the RMS), the corresponding energy and charge jitter of the electron beams are estimated to be 0–3.6% and 17.4%, respectively. Despite these fluctuations, the FEL remains operable in the saturation regime, as demonstrated by the stability analysis detailed in Ref. [78].

The tailored density profile in our scheme is based on the configuration demonstrated in previous experiments [19, 44], which employs a perforated baffle upstream of a supersonic helium nozzle. By adjusting the baffle–nozzle separation and backpressure, a range of density distributions can be readily obtained. Alternative approaches have also been investigated, including those based on a supersonic nozzle with a thin metal blade [27, 79, 80], modifications to the metal plate structure [81], and optical field ionization [82]. These methods yield shock front lengths from tens to hundreds of micrometers, covering the parameter space of the presented scheme.

## V. CONCLUSIONS

In conclusion, we have developed a single-stage configuration for generating high-quality *e*-beams in LWFAs using the synergetic injection mechanism. The source of the beam quality degradation has been identified, and targeted optimization strategies have been developed, focusing on the global (slice) energy spread and the projected (slice) normalized emittance. Under optimized conditions, the RMS global (average slice) energy spread can be compressed to 0.9% (~0.2%), while the projected (average slice) normalized emittance can reach 0.094 (0.036) and 0.301 (0.106) mm mrad in the horizontal and vertical directions, respectively. As an example, we demonstrate the effectiveness of our proposed scheme for driving compact FELs using a start-to-end model. The resulting FEL radiation achieves saturation in a 4.5-meter-long undulator, with an energy of 17.4 μJ, a power of 6.0 GW and a central wavelength of 23.9 nm. The presented scheme not only provides a framework for optimizing beam quality in LWFAs, but also offers a pathway to achieve high-gain FEL driven by LWFA at the laboratory scale, paving the way for compact and widely accessible systems suitable for user-oriented applications.

## ACKNOWLEDGMENTS

This work was supported by the National Natural Science Foundation of China (Grant No. 12388102), the Strategic Priority Research Program of the Chinese Academy of Sciences (Grant No. XDB0890200), the National Natural Science Foundation of China (Grant Nos. 12225411 and 12474349), the National Key R&D Program of China (Grant No. 2025YFF0515000), CAS Project for Young Scientists in Basic Research (Grant No. YSBR060), CAS Youth Innovation Promotion Association (No. 2022242), the Ningbo Yongjiang Talent Programme (Grant No. 2025A-104-G), Zhangjiang Laboratory, and the New Cornerstone Science Foundation through the XPLORER PRIZE.

## DATA AVAILABILITY

The data that support the findings of this article are openly available [83].